\documentclass[a4paper,3p,twocolumn,preprint]{elsarticle}

 \journal{Physica D}

\usepackage{amsmath}
\usepackage{amssymb}
\usepackage{graphicx}
\usepackage{bm}
\usepackage{epstopdf}

\newcommand{\la}{\lambda}

\newcommand{\om}{\omega}

\newcommand{\prt}{\partial}

\begin{document}

\begin{frontmatter}

\title{
Quasiclassical integrability condition in AKNS scheme}

\author{A.~M.~Kamchatnov}
\ead{kamch@isan.troitsk.ru}
\author{D.~V.~Shaykin}
\ead{shaykin.dv@phystech.edu}
\address{
Institute of Spectroscopy, Russian Academy of Sciences, Troitsk,
Moscow, 142190, Russia }
\address{Moscow Institute of Physics and Technology, Institutsky lane 9, Dolgoprudny, Moscow region, 141700, Russia}
\address{
Skolkovo Institute of Science and Technology, Skolkovo, Moscow, 143026, Russia} 

\date{\today}

\begin{abstract}
In this paper, we study the condition of quasiclassical integrability of soliton equations.
This condition states that the Hamiltonian structure of equations, which govern propagation
of high-frequency wave packets, is preserved by the dispersionless flow independently of
initial conditions. If this condition is fulfilled, then the carrier wave number of any
packet is a certain function of local values of the dispersionless variables pertained to the
soliton equations under consideration. We show for several examples that this function together 
with the dispersion relation for linear harmonic waves determine the quasiclassical limit of 
the Lax pair functions in the scalar representation of the Ablowitz-Kaup-Newell-Segur scheme.
\end{abstract}



\end{frontmatter}


\section{Introduction}

The general definition of the ``integrability'' property of nonlinear wave equations is apparently
impossible (see, e.g., discussion in Ref.~\cite{fnt-1991}). In a narrow sense, integrability is often
related with existence of the Lax pair of linear systems whose compatibility condition leads to the
nonlinear wave equations under consideration. First discovered for the Korteweg-de Vries (KdV)
equation \cite{ggkm-67,lax-68} and nonlinear Schr\"{o}dinger (NLS) equation \cite{zs-71}, Lax pairs
were found for a number of wave equations and they were widely used in various physical applications
(see, e.g., Refs.~\cite{lam-80,dj-89,scott-03} and reference therein). However, the relationship between
the physical properties of wave equations and associated with them Lax pairs still remains unclear.

As was noticed in Ref.~\cite{sk-23} in the theory of the generalized NLS equation
\begin{equation}\label{eq1}
i\psi_t+\frac{1}{2}\psi_{xx} - f(|\psi|^2)\psi = 0,
\end{equation}
the integrable case with $f(|\psi|^2)=|\psi|^2$ is distinguished by a very special property of
propagation of high-frequency wave packets along large scale background waves, and this property
can be formulated in purely physical characteristics of Eq.~(\ref{eq1}). In the problem of packets
propagation we have two very different characteristic length parameters: a small wavelength
$\sim k^{-1}$ of harmonics that compose the packet ($k$ is the carrier wave number) and the size $\sim l$
of the background wave, so $(kl)^{-1}$ is a small parameter of the theory. Its existence allows one
to separate dispersionless evolution of a background wave from propagation of a linear wave
packet. The dispersionless evolution is described by the hydrodynamic equations
\begin{equation}\label{eq2}
\begin{split}
\rho_t + (\rho u)_x = 0, \quad
u_t + uu_x + \frac{c^2}{\rho} \rho_x  = 0,
\end{split}
\end{equation}
where $\rho$ and $u$ are defined according to the formula
\begin{equation}\label{eq2a}
\psi(x,t) = \sqrt{\rho(x,t)}\exp\left(i\int^x u(x',t)dx'\right)
\end{equation}
and
\begin{equation}\label{eq3}
  c^2=\rho f'(\rho).
\end{equation}
The small amplitude wave packet propagates along smooth distributions $\rho=\rho(x,t), u=u(x,t)$,
given by some specific solution of Eqs.~(\ref{eq2}), according to the Bogoliubov dispersion relation
\begin{equation}\label{eq4}
\omega = k\left( u \pm \sqrt{c^2+\frac{k^2}{4}} \right),
\end{equation}
where due to the condition $kl\gg1$ one can consider $u$ and $c=c(\rho)$ constant within the packet's
width. This means that the packet is considered as a point-like particle with coordinate $x=x(t)$
whose motion obeys the Hamilton equations (see, e.g., \cite{synge-37,ko-90})
\begin{equation}\label{eq5}
\frac{d x}{d t} = \frac{\prt \omega }{\prt k},\qquad
\frac{d k}{d t} = -\frac{\prt \omega }{\prt x}.
\end{equation}
As we showed in Ref.~\cite{sk-23}, the system (\ref{eq2}), (\ref{eq5}) admits the solution
in the form $k=k(\rho,u)$, where the wave number depends on $(x,t)$-coordinates via the
local values $\rho$ and $u$ of the background wave, only in case of the integrable NLS
equation with $f(\rho)=\rho$. In this case we obtain
\begin{equation}\label{eq6}
  k^2=(q-u)^2-4\rho,
\end{equation}
where $q$ is an integration constant. As was shown in Ref.~\cite{kamch-23}, the
expression in the right-hand side of this formula is related with the
quasiclassical limit of the Zakharov-Shabat spectral problem \cite{zs-71}
($q$ plays the role of the spectral parameter), whereas the quasiclassical limit of
the other equation of the corresponding Lax pair reduces to the number of waves
conservation law (see \cite{whitham-65,whitham})
\begin{equation}\label{eq7}
  k_t+\om_x=0.
\end{equation}

Thus, both equations of the Lax pair in their quasiclassical limit for the NLS equation turn out
to be related with dispersion relation $\om=\om(k)$ of linear waves and nonlinearity function
$f(\rho)=\rho$ which appears in the last term of the right-hand side of Eq.~(\ref{eq6}). 
Substitution of this formula into the first Hamilton equation (\ref{eq5}) yields very simple 
equation for propagation of wave packets,
\begin{equation}\label{eq8}
  \frac{dx}{dt}=q-\frac{2\rho(x,t)}{q-u(x,t)},
\end{equation}
where $q$ is determined by the initial condition $k(x=x_0,t=t_0)=k_0$ for Eq.~(\ref{eq6}).

In this paper, we formulate the observation of Ref.~\cite{kamch-23} in more general terms and
confirm validity of such a general formulation by some other examples of  integrable equations
which belong to the Ablowitz-Kaup-Newell-Segur (AKNS) scheme \cite{akns-74}. We derive for 
these examples the relations similar
to Eq.~(\ref{eq6}), demonstrate their connection with the
quasiclassical limit of the corresponding Lax pairs, and obtain equations of motion of
high-frequency wave packets along non-uniform and time dependent background waves of the
dispersionless limit.

\section{General theory}

We assume that a physical system is characterized by two variables which in many situations
have physical sense of ``density'' $\rho(x,t)$ and ``flow velocity'' $u(x,t)$. The two
natural limits can be distinguished in dynamics of such systems. First, we can consider
infinitely small amplitude waves propagating along uniform stationary background $\rho=\rho_0$,
$u=u_0$. Then we obtain the dispersion relation for harmonic waves
$\rho-\rho_0,u-u_0\propto\exp[i(kx-\om t)]$,
\begin{equation}\label{eq9}
  \om=\om(k,\rho_0,u_0),
\end{equation}
(see, e.g., Eq.~(\ref{eq4})). This dispersion relation often includes some parameter $k_D$,
so that Eq.~(\ref{eq9}) becomes linear for $k\ll k_D$,
\begin{equation}\label{eq10}
  \om\approx(u_0\pm c_0)k,\qquad k\ll k_D,
\end{equation}
where $c_0$ is called ``sound velocity'' and Eq.~(\ref{eq10}) corresponds to long sound waves
propagating upstream or downstream the flow moving with velocity $u_0$.

As another typical limit, we can consider long waves that change little on distances about 
$\sim k_D^{-1}$
(i.e., $|\prt\rho/\prt x|\sim\rho/l$,  $|\prt u/\prt x|\sim u/l$, and $lk_D\gg1$). In this
case we arrive at the so-called ``dispersionless'' or ``hydrodynamic'' limit of our
equations (see, e.g., Eqs.~(\ref{eq2})). Generally speaking, in case of two wave variables
the hydrodynamic equations can be transformed to the diagonal form
\begin{equation}\label{eq11}
\begin{split}
  &\frac{\prt r_+}{\prt t}+v_+(r_+,r_-)\frac{\prt r_+}{\prt x}=0,\\
  &\frac{\prt r_-}{\prt t}+v_-(r_+,r_-)\frac{\prt r_-}{\prt x}=0
   \end{split}
\end{equation}
for ``Riemann invariants'' $r_{\pm}=r_{\pm}(\rho,u)$, where the characteristic velocities are
given by the formulas
\begin{equation}\label{eq12}
  v_{\pm}=u\pm c
\end{equation}
for sound waves propagation in the long wavelength limit (\ref{eq10}) and they can also be
expressed as functions of the Riemann invariants, $v_{\pm}=v_{\pm}(r_+,r_-)$.

Now, if we consider a high frequency wave packet with carrier wave number $k$ and width $d$
such that
\begin{equation}\label{eq13}
  2\pi/k\ll d\ll l,
\end{equation}
then we can define the packet's coordinate $x(t)$ with good enough accuracy. As is well known,
such a packet propagates with the group velocity and its local wavelength changes due to
refraction (dependence of the phase velocity $\om/k$ on the coordinate $x$), so the packet's
dynamics obeys the Hamilton equations (\ref{eq5}) with
\begin{equation}\label{eq14}
  \om=\om(k,r_+,r_-).
\end{equation}
As was shown in Ref.~\cite{sk-23}, if we demand that the Hamiltonian form of Eqs.~(\ref{eq5})
is preserved by the hydrodynamic flow (\ref{eq11}), then the carrier wave number $k$ must
only depend on the local values of the dispersionless variables $r_+,r_-$,
\begin{equation}\label{eq15}
  k=k(r_+,r_-,q),
\end{equation}
which satisfied the equations
\begin{equation}\label{eq16}
  \frac{\prt k}{\prt r_+}=\frac{\prt\om/\prt r_+}{v_+-\prt\om/\prt k},\quad
  \frac{\prt k}{\prt r_-}=\frac{\prt\om/\prt r_-}{v_--\prt\om/\prt k},
\end{equation}
and $q$ is an integration constant.
Naturally, solution of these equations does only exist if
\begin{equation}\label{eq17}
  \frac{\prt}{\prt r_+}\left(\frac{\prt k}{\prt r_-}\right)=
  \frac{\prt}{\prt r_-}\left(\frac{\prt k}{\prt r_+}\right).
\end{equation}
Usually this condition is only fulfilled in the limit of large $k$, so then one can obtain
an approximate solution which is enough for discussion of propagation of high frequency
wave packets. However, as was noticed in Ref.~\cite{sk-23}, Eq.~(\ref{eq17}) is satisfied
identically for the integrable case of Eq.~(\ref{eq1}) with $f(\rho)=\rho$, so that the
exact solution of Eqs.~(\ref{eq16}) can be found (see Eq.~(\ref{eq6}) with $\rho$ and $u$
expressed in terms of the Riemann invariants). We will show below that the same statement
turns out to be true for other examples of integrable equations in AKNS scheme.

The above theory refers to any wave packet. Let us consider now a particular case of wave
packets corresponding to the small amplitude edges of dispersive shock waves propagating
along smooth parts of large wave pulses. As was noticed in Ref.~\cite{gp-87}, oscillations
enter into the dispersive shock region with velocity equal to difference between the group and
phase velocities of this small-amplitude edge wave packet. In situations when these
oscillations transform eventually into separate solitons, this remark allows one to write
the formula for the number of solitons as (see Refs.~\cite{kamch-21a,kamch-20a})
\begin{equation}\label{eq18}
  N=\frac1{2\pi}\int_0^{\infty}\left(k\frac{\prt\om}{\prt k}-\om\right)dt=\frac{S}{2\pi},
\end{equation}
where $S$ is the action produced by a point-like particle moving according to the
Hamilton equations (\ref{eq5}) with Hamiltonian $\om$. The integral here is taken over
the whole packet's path $x=x(t),k=k(t)$ from the wave breaking moment $t=0$ till large
enough time $t\to\infty$ of full transformation of the pulse into solitons. We can also
take the path back with $k$ replaced by $-k$, so the direct and inverse paths are combined
to a closed contour $C$ in the phase space $(x,k)$ and Eq.~(\ref{eq18}) can be written in
the form of the Poincar\'e-Cartan integral invariant (see Ref.~\cite{kamch-23} and
references therein)
\begin{equation}\label{eq19}
  N=\frac1{\pi}\oint(k\delta x-\om\delta t),
\end{equation}
where $\delta x=(\prt\om/\prt k)\delta t$. As was shown in Ref.~\cite{kamch-23}, the
Poincar\'e-Cartan integral invariant is preserved by the hydrodynamic flow (\ref{eq11})
for $k$ given by solution (\ref{eq15}) of Eqs.~(\ref{eq16}). Consequently, we can transform
the contour $C$ to the initial state $t=0$ and then Eq.~(\ref{eq19}) gives the expression
\begin{equation}\label{eq20}
  N=\frac{1}{2\pi}\int_{-\infty}^{\infty}k(r_+^0(x),r_-^0(x))dx,
\end{equation}
where $r_{\pm}^0(x)$ are the initial distributions of the dispersionless Riemann invariants.
For a particular case of simple wave distributions, when one of the Riemann invariants is
constant, such a formula was obtained by a different method in
Refs.~\cite{egkkk-07,egs-08,mfweh-20} and by direct calculation of the integral in
Eq.~(\ref{eq18}) in Refs.~\cite{kamch-20a,kamch-21,cbk-21}.

Now we will relate Eq.~(\ref{eq20}) with the Lax pair in AKNS scheme. It is convenient to
discuss its quasiclassical limit in scalar representation \cite{ak-02}, when the integrable
equation under consideration is expressed as a compatibility condition of linear equations
\begin{equation}\label{eq21}
  \phi_{xx}=\mathcal{A}\phi,\qquad \phi_t=-\frac12  \mathcal{B}_x\phi+ \mathcal{B}\phi_x.
\end{equation}
We will give concrete examples of this scheme in the next Section and here we will consider
some its general properties. To this end, we notice that the second order differential
equation (\ref{eq21}) has two basis solutions $\phi_+,\phi_-$, so we define the
`squared basis function',
\begin{equation}\label{eq22}
  g=\phi_+\phi_-,
\end{equation}
which satisfies the equations
\begin{equation}\label{eq23}
 g_{xxx}-2\mathcal{A}_x{g}-4\mathcal{A}{g}_x=0, \qquad
{g}_t=\mathcal{B}{g}_x-\mathcal{B}_x{g}.
\end{equation}
The first equation can be integrated to give
\begin{equation}\label{eq24}
\frac12{g}{g}_{xx}-\frac14{g}_x^2-\mathcal{A}{g}^2={P},
\end{equation}
where $P$ is an integration constant. Here $\mathcal{A}$ depends, besides the wave variables,
on the spectral parameter $\la$ and we normalize $\mathcal{A}$ in such a way that
\begin{equation}\label{eq25}
  \mathcal{A}\to-\sigma\la^r\quad\text{as}\quad \la\to\infty,
\end{equation}
and $\sigma=\pm1$ denotes the sign of this leading term in this series expansion of $\mathcal{A}$,
$r$ being the exponent of the leading term. Then in case of polynomial dependence of $g$ on
$\la$ we get
\begin{equation}\label{eq26}
  \frac{\sqrt{P/\la^r}}{g}\to 1 \quad\text{as}\quad \la\to\infty.
\end{equation}
The second equation (\ref{eq23}) yields the generating function of conservation laws
\begin{equation}\label{eq27}
\left(\frac{\sqrt{P}}{{g}}\right)_t-
\left(\frac{\sqrt{P}}{{g}}\mathcal{B}\right)_x=0,
\end{equation}
where inessential factor $\la^{-r/2}$ is dropped out and the expressions $\sqrt{P}/g$ and
$\sqrt{P}\mathcal{B}/g$ can be expanded is series in powers of $\la^{-1}$ to give a series of
particular conservation laws. This makes Eq.~(\ref{eq27}) a convenient tool for derivation
of the Whitham modulation equations \cite{kamch-94,kamch-04}.

Solutions of Eqs.~(\ref{eq21}) can also be expressed in terms of $g$ \cite{ak-01}
\begin{equation}\label{eq28}
  \phi_{\pm}=\sqrt{g}\exp\left(\pm i\int^x\frac{\sqrt{P}}{g}\,dx\right).
\end{equation}
The Whitham averaging method in Krichever's formulation \cite{krichever-88} assumes that
Eq.~(\ref{eq28}) remains correct during the whole process of transformation of the
initially smooth pulse to a trains of separate solitons. At the initial moment of time
the wave variables are smooth functions of $x$ and they can be approximated by the
dispersionless Riemann invariants $r_{\pm}^0$. Then the function $g$ is also smooth and
the terms with its derivatives in Eq.~(\ref{eq24}) can be neglected, so we arrive at the
approximate expression
\begin{equation}\label{eq29}
  \frac{\sqrt{P}}{g}\approx\sqrt{-\sigma\overline{\mathcal{A}}}\equiv\overline{k}(r_+^0,r_-^0,\la),
\end{equation}
where $\overline{\mathcal{A}}=\overline{\mathcal{A}}(r_+^0,r_-^0,\la)$ is to be obtained from
$\mathcal{A}$ in the same approximation with omitted derivatives of $r_+^0$ and $r_-^0$ 
(or $\rho$ and $u$). Consequently, we obtain a quasiclassical limit of Eq.~(\ref{eq28}),
\begin{equation}\label{eq30}
  \phi_{\pm}\approx\sqrt{g}\exp\left(\pm i\int^x\overline{k}(r_+^0,r_-^0,\la)\,dx\right).
\end{equation}
The condition that $\phi_{\pm}$ are single-valued functions of $x$ gives the Bohr-Sommerfeld
quantization rule (see, e.g., \cite{karpman-67,karpman-73,JLML-99,kku-02}) and, in particular,
the formula for the number of solitons
\begin{equation}\label{eq31}
\begin{split}
  \int_{-\infty}^{\infty}\overline{k}(r_+^0(x),r_-^0(x),\la_n)\,dx=\pi N,
     \end{split}
\end{equation}
where $N$ is the largest quantum number equal to the number of solitons and we replaced the
corresponding turning points $x_1^N,x_2^N$ by $\mp\infty$, since contribution of the intervals
$-\infty<x<x_1^N, x_2^N<x<\infty$ is supposed to be negligibly small in this quasiclassical
approximation. Comparison of Eqs.~(\ref{eq20}) and (\ref{eq31}) gives the expression
\begin{equation}\label{eq32}
  k(r_+^0,r_-^0,q)=2\overline{k}(r_+^0,r_-^0,\la),
\end{equation}
which defines the relationship
\begin{equation}\label{eq33}
  \la=\la(q)
\end{equation}
between the integration constant $q$ in (\ref{eq15}) and the spectral parameter $\la$ in the
Lax pair (\ref{eq21}).

Substitution of $\sqrt{P}/g=k/2$ into Eq.~(\ref{eq27}) gives
\begin{equation}\label{eq34}
  k_t-(k\overline{\mathcal{B}})_x=0,
\end{equation}
where $\overline{\mathcal{B}}(r_+^0,r_-^0,\la)$ is obtained from $\mathcal{B}$ in the same
quasiclassical approximation with neglected $x$-derivatives of the Riemann invariants.
This equation must coincide with Eq.~(\ref{eq7}), so we get
\begin{equation}\label{eq35}
  \frac{\om}{k}=-\overline{\mathcal{B}}.
\end{equation}
Thus, we arrive at the following expressions for the quasiclassical limits of $\mathcal{A}$
and $\mathcal{B}$:
\begin{equation}\label{eq36}
\begin{split}
  \overline{\mathcal{A}}(r_+,r_-,\la(q))&=-\frac14k^2(r_+,r_-,q),\\
  \overline{\mathcal{B}}(r_+,r_-,\la)q))&=-\frac{\om(k(r_+,r_-,q),r_+,r_-)}{k(r_+,r_-,q)},
  \end{split}
\end{equation}
where $k^2$ is defined by solution of Eqs.~(\ref{eq16}). The condition (\ref{eq17}) of
existence of such a solution becomes the condition of existence of the functions $\overline{\mathcal{A}}$
and $\overline{\mathcal{B}}$, that is it plays the role of the `integrability test' for the equation
under consideration in framework of AKNS scheme at least in the quasiclassical limit.

Validity of the above expressions was illustrated in Ref.~\cite{kamch-23} by an example
of NLS equation. Here we will consider several other examples.

\section{Examples}

\subsection{Kaup-Boussinesq system}

At first we will consider the Kaup-Boussinesq system
\begin{equation}\label{eq37}
  h_t+(hu)_x-\frac14h_{xxx}=0,\quad u_t+uu_x+h_x=0,
\end{equation}
which describes propagation of waves on shallow water \cite{bouss-1877,kaup-75} and in
two-component Bose-Einstein condensates \cite{ikcp-17}. In the shallow-water context,
$h(x,t)$ is the local depth of water and $u(x,t)$ is its mean horizontal flow velocity.
Linearized equations lead to the dispersion relation for harmonic waves,
\begin{equation}\label{eq38}
  \om(k,h,u)=k\left(u\pm\sqrt{h+\frac{k^2}{4}}\right).
\end{equation}
Equations of the dispersionless limit
\begin{equation}\label{eq39}
  h_t+(hu)_x=0,\quad u_t+uu_x+h_x=0
\end{equation}
are cast to the diagonal form (\ref{eq11}) by introduction of the Riemann invariants
\begin{equation}\label{eq40}
  r_{\pm}=\frac{u}{2}\pm\sqrt{h},
\end{equation}
so that
\begin{equation}\label{eq41}
  v_+=\frac12(3r_++r_-),\qquad v_-=\frac12(r_++3r_-)
\end{equation}
and
\begin{equation}\label{eq42}
  u=r_++r_-,\qquad h=\frac14(r_+-r_-)^2.
\end{equation}
It is convenient to transform Eqs.~(\ref{eq16}) to independent variables $h$ and $u$
instead of $r_+$ and $r_-$. Straightforward calculation gives
\begin{equation}\label{eq43}
  \frac{\prt k^2}{\prt h}=-4,\qquad \frac{\prt k^2}{\prt u}=-2\sqrt{k^2+4h},
\end{equation}
and we get the solution in the form
\begin{equation}\label{eq44}
  k^2=(q-u)^2-4h,
\end{equation}
$q$ is an integration constant. Consequently, Eqs.~(\ref{eq36}) give the quasiclassical
limit of the Lax pair functions (we assume $q>u$ and the upper sign in the dispersion
relation (\ref{eq38}))
\begin{equation}\label{eq45}
  \overline{\mathcal{A}}=h-\left(\frac{q}2-\frac{u}2\right)^2,\quad
  \overline{\mathcal{B}}=-\left(\frac{q}2+\frac{u}2\right).
\end{equation}
It is remarkable that in this case the quasiclassical limit coincides with the exact
expressions \cite{kaup-75} for the functions $\mathcal{A}$ and $\mathcal{B}$, if we
identify $q/2$ with the spectral parameter $\la$.

The theory of propagation of high-frequency wave packets actually coincides with that
theory for the defocusing NLS equation \cite{sk-23}, since the dispersion relation
(\ref{eq38}) and dispersionless equations (\ref{eq39}) are mathematically identical.
Therefore substitution of Eq.~(\ref{eq44}) into the first Hamilton equation (\ref{eq5})
gives Eq.~(\ref{eq8}) with $\rho(x,t)$ replaced by $h(x,t)$.

Eqs.~(\ref{eq38}) and (\ref{eq39}) satisfy the `integrability test', that is the
derivatives (\ref{eq43}) commute and define the same function (\ref{eq44}) as
(\ref{eq6}) in the NLS equation case, but we have two different dispersive generalizations
with different Lax pairs. It is worth noticing that these two integrable systems, NLS and
Kaup-Boussinesq equations, can be obtained as two different approximations to the
integrable Landau-Lifshitz equations with an easy-plane anisotropy \cite{ikcp-17}.

\subsection{Derivative nonlinear Schr\"{o}dinger equation}

Derivative nonlinear  Schr\"{o}dinger (DNLS) equation
\begin{equation}\label{eq46}
  i\psi_t+\frac12\psi_{xx}-i(|\psi|^2\psi)_x=0
\end{equation}
plays an important role in the theory of nonlinear Alfv\'{e}n waves (see, e.g.,
Ref.~\cite{kbhp-88}). It is transformed by means of the substitution (\ref{eq2a})
to a hydrodynamic-like form
\begin{equation}\label{eq48}
  \begin{split}
  & \rho_t+\left[\rho\left(u-\frac32\rho\right)\right]_x=0,\\
  &u_t+uu_x-(\rho u)_x+\left(\frac{\rho_x^2}{8\rho^2}-\frac{\rho_{xx}}{4\rho}\right)_x=0.
  \end{split}
\end{equation}
Linearization of these equations yields the dispersion relation for harmonic waves
\begin{equation}\label{eq49}
  \om=k\left(u-2\rho\pm\sqrt{\rho(\rho-u)+\frac{k^2}{4}}\right).
\end{equation}
The uniform state is modulationally stable for $\rho>u$. Then the dispersionless limit
equations
\begin{equation}\label{eq50}
  \begin{split}
  & \rho_t+\left[\rho\left(u-\frac32\rho\right)\right]_x=0,\\
  &u_t+uu_x-(\rho u)_x=0
  \end{split}
\end{equation}
have real characteristic velocities
\begin{equation}\label{eq51}
  v_{\pm}=u-2\rho\pm\sqrt{\rho(\rho-u)}
\end{equation}
and can be transformed to a diagonal form (\ref{eq11}) by introduction of the Riemann invariants
\begin{equation}\label{eq52}
  r_{\pm}=\frac{u}{2}-\rho\pm\sqrt{\rho(\rho-u)}.
\end{equation}
The velocities (\ref{eq51}) are expressed in terms of the Riemann invariants by Eqs.~(\ref{eq41}).
Again we transform Eqs.~(\ref{eq16}) to independent variables $\rho,u$ and obtain
\begin{equation}\label{eq53}
  \begin{split}
  &\frac{\prt k^2}{\prt\rho}=4\left[\sqrt{k^2+4\rho(\rho-u)}-(2\rho-u)\right],\\
  &\frac{\prt k^2}{\prt u}=-2\left[\sqrt{k^2+4\rho(\rho-u)}-2\rho\right].
  \end{split}
\end{equation}
These derivatives commute and admit the solution
\begin{equation}\label{eq54}
\begin{split}
  k^2 &=(q-u)^2+4q\rho\\
  &\equiv (2\rho-u+q)^2-4\rho(\rho-u),
  \end{split}
\end{equation}
$q$ is an integration constant.
Assuming $2\rho-u+q>0$ and taking the upper sign in Eq.~(\ref{eq49}), we obtain from Eqs.~(\ref{eq36})
\begin{equation}\label{eq55}
  \overline{\mathcal{A}}=-\frac14(q-u)^2-q\rho,\quad
  \overline{\mathcal{B}}=-\frac{q}2+\frac{u}2+\rho.
\end{equation}

The Lax pair for the DNLS equation was found in Ref.~\cite{kn-78} and in the scalar form
it reads \cite{ak-02}
\begin{equation}\label{eq56}
  \begin{split}
  &\mathcal{A}=-\frac14\left(\la^2-\frac{i\psi_x}{\psi}\right)^2+4\la^2\rho
  -\left(\frac{i\psi_x}{2\psi}\right)_x,\\
  &\mathcal{B}=2\la^2+\rho+\frac{i\psi_x}{2\psi}.
  \end{split}
\end{equation}
After substitution of Eq.~(\ref{eq2a}) and taking the dispersionless limit of negligibly small
derivatives of $\rho$ and $u$, we get
\begin{equation}\label{eq57}
  \overline{\mathcal{A}}=-\frac14(\la^2+u)^2+4\la^2\rho,\quad
  \overline{\mathcal{B}}=2\la^2-\frac{u}2+\rho.
\end{equation}
These formulas coincide with Eqs.~(\ref{eq55}) for $q=-4\la^2$.

Substitution of Eq.~(\ref{eq54}) into the first Hamilton equation (\ref{eq5}) gives a convenient
form of equation of motion for high-frequency wave packets
\begin{equation}\label{eq58}
  \frac{dx}{dt}=q-\frac{2\rho(\rho-u)}{2\rho-u+q},
\end{equation}
where $\rho=\rho(x,t), u=u(x,t)$ are given by some solution of the dispersionless
Eqs.~(\ref{eq50}) and $q$ is determined by the initial conditions.

\subsection{Landau-Lifshitz equation}

Landau-Lifshitz equation
\begin{equation}\label{eq59}
  \mathbf{M}_t=(\mathbf{M}_{xx}-M_3\mathbf{e}_3)\wedge\mathbf{M}
\end{equation}
for the magnetization vector $\mathbf{M}=(M_1,M_2,M_3)$, $|\mathbf{M}|=1$, with an easy-plane
anisotropy ($\mathbf{e}_3$ is a unit vector of the $z$ axis normal to the plane of anisotropy)
can be transformed to more convenient variables $w,v$ according to the formulas
\cite{ikcp-17,ckp-16,ish-17}
\begin{equation}\label{eq60}
\begin{split}
  M_3&=-w,\\
  M_-&\equiv M_1-iM_2\\
  &=\sqrt{1-w^2}\exp\left(-i\int^xvdx'\right),
  \end{split}
\end{equation}
so we get
\begin{equation}\label{eq61}
  \begin{split}
   w_t&-\left[v(1-w^2)\right]_x=0,\\
   v_t&-\left[w(1-v^2)\right]_x\\
  &+\left[\frac{1}{\sqrt{1-w^2}}\left(\frac{w_x}{\sqrt{1-w^2}}\right)_x\right]_x=0.
  \end{split}
\end{equation}
As usual, we obtain the dispersion relation
\begin{equation}\label{eq62}
  \om=k\left(2wv\pm\sqrt{k^2+(1-v^2)(1-w^2)}\right)
\end{equation}
for harmonic waves, that is we get the condition $|v|\leq1$ of modulation stability
of a uniform state (the condition $|w|\leq1$ is satisfied by definition). The
dispersionless equations
\begin{equation}\label{eq63}
  \begin{split}
   w_t&-\left[v(1-w^2)\right]_x=0,\\
   v_t&-\left[w(1-v^2)\right]_x=0
   \end{split}
\end{equation}
have characteristic velocities
\begin{equation}\label{eq64}
  v_{\pm}=2v w\pm\sqrt{(1-v^2)(1-w^2)}
\end{equation}
and can be diagonalized by introduction of the Riemann invariants
\begin{equation}\label{eq65}
  r_{\pm}=v w\pm\sqrt{(1-v^2)(1-w^2)}.
\end{equation}
As a result, Eqs.~(\ref{eq63}) are cast to the form (\ref{eq11}) with
$v_{\pm}$ given by Eqs.~(\ref{eq41}). Eqs.~(\ref{eq16}) transformed to independent
variables $w,v$ have simple symmetrical form
\begin{equation}\label{eq66}
  \begin{split}
  & \frac{\prt k^2}{\prt w}=2\left[w(1-v^2)-v\sqrt{k^2+(1-v^2)(1-w^2)}\right],\\
  & \frac{\prt k^2}{\prt v}=2\left[v(1-w^2)-w\sqrt{k^2+(1-v^2)(1-w^2)}\right].
  \end{split}
\end{equation}
These derivatives commute and yield the solution
\begin{equation}\label{eq67}
  k^2=(q-v w)^2-(1-v^2)(1-w^2).
\end{equation}
Assuming $q-vw>0$, we obtain the expressions
\begin{equation}\label{eq68}
\begin{split}
  \overline{\mathcal{A}}&=\frac14\left[(1-v^2)(1-w^2)-(q-v w)^2\right],\\
  \overline{\mathcal{B}}&=-q-vw.
  \end{split}
\end{equation}

The Lax pair for the Landau-Lifshitz equation was found in Ref.~\cite{br-81} and in a scalar
representation it takes the form \cite{ak-02}
\begin{equation}\label{eq69}
  \begin{split}
  \mathcal{A}&=\frac14\left(i\la M_3-\frac{(M_-)_x}{M_-}\right)^2+\frac14(1-\la^2)(1-M_3^2)\\
  &+\frac12\left(i\la M_3-\frac{(M_-)_x}{M_-}\right)_x,\\
  \mathcal{B}&=-\la-i\left[(M_3)_x-M_3\frac{(M_-)_x}{M_-}\right].
  \end{split}
\end{equation}
Substitution of Eqs.~(\ref{eq60}) and taking the dispersionless limit easily reproduce
Eqs.~(\ref{eq68}) with $\la=q$.

At last, substitution of Eq.~(\ref{eq67}) into the expression for the group velocity
yields the equation of motion for high-frequency wave packets
\begin{equation}\label{eq70}
  \frac{dx}{dt}=2q-\frac{(1-v^2)(1-w^2)}{q-vw},
\end{equation}
where $v=v(x,t),w=w(x,t)$ are given by some solution of Eqs.~(\ref{eq63}) and $q$ is defined by
initial conditions.

\section{Conclusion}

We showed in this paper that the condition of quasiclassical integrability leads to important
relationship between the carrier wave number $k$ of high-frequency wave packets propagating
along smooth background waves and the local values $(r_+,r_-)$ of the background variables.
This relationship does not depend on initial conditions for the background wave evolution.
We showed that the function $k=k(r_+,r_-)$ and the dispersion relation
$\om=\om(k,r_+,r_-)$ are related with the functions $\overline{\mathcal{A}},\overline{\mathcal{B}}$
which represent the Lax pairs for the corresponding equations in AKNS scheme in
quasiclassical limit.

The obtained here relationship $k=k(r_+,r_-)$ has several important applications. Its substitution 
into expression for the group velocity yields a convenient equation of motion for wave packets.
On the other side, an analytical continuation of this formula on complex values of $k=i\kappa$
gives the expression for the inverse half-width $\kappa$ of solitons as a function of the
local dispersionless variables, so we obtain the soliton's equation of motion along a smooth
background wave. Besides that, the function $k(r_+,r_-)$ is used in formulation of the
Bohr-Sommerfeld quantization rule for parameters of solitons emerging from an intensive
initial pulse. At last, if the equation under consideration is not completely integrable,
but Eqs.~(\ref{eq16}) admit an approximate solution in the limit of large $k$, then we can 
apply it to non-integrable equations on the same footing as for integrable ones. Some
examples of such an extended theory were considered in Refs.~\cite{sk-23,kamch-23,ks-23}.
We believe that such an approach will find many other applications.

\section*{Acknowledgments}

This research is funded by the research project FFUU-2021-0003 of the Institute of Spectroscopy
of the Russian Academy of Sciences (Section~II) and by the RSF grant number~19-72-30028
(Section~III).


\begin{thebibliography}{99}

\bibitem{fnt-1991} H. Flaschka, A. C. Newell and M. Tabor, Integrability, in
{\it What Is Integrablity?} Ed. V.~E.~Zakharov, p.~73, Springer, Berlin, 1991.

\bibitem{ggkm-67} C. S. Gardner, J. M. Green, M. D. Kruskal and R. M. Miura, Phys. Rev. Lett., 19 (1967) 1095.

\bibitem{lax-68} P. D. Lax, Commun. Pure Appl. Math., 21 (1968) 467.

\bibitem{zs-71} V. E. Zakharov and A. B. Shabat, Zh. Eksp. Teor. Fiz., 61 (1971) 118;
Sov. Phys. JETP, 34 (1972) 62.

\bibitem{lam-80} G. L. Lamb, jr., Elements of Soliton Theory, Wiley, N. y., 1980.

\bibitem{dj-89} P. G. Drazin and R. S. Johnson, Solitons: An Introduction, CUP, Cambridge, 1989.

\bibitem{scott-03} A. Scott, Nonlinear Science. Emergemce and Dynamics of Coherent Sctructures,
Oxford University Press, Oxford, 2003.

\bibitem{sk-23} D. V. Shaykin and A. M. Kamchatnov,  Phys. Fluids, 35 (2023) 062108.

\bibitem{synge-37} J.~L.~Synge, { Geometrical Optics. An Introduction into Hamilton's method,}
CUP, Cambridge, 1937.

\bibitem{ko-90} Yu.~A.~Kravtsov, Yu.~I.~Orlov, { Geometrical Optics of Inhomogeneous Media,}
Springer, Berlin, 1990.

\bibitem{kamch-23} A. M. Kamchatnov, Asymptotic theory of not completely integrable soliton equations,
preprint arxiv:2305.12346 (2023).

\bibitem{whitham-65} G. B. Whitham, Proc. Roy. Soc. Lond., A 283 (1965) 238.

\bibitem{whitham} G. B. Whitham, { Linear and Nonlinear Waves,} Wiley, New York, 1974.

\bibitem{akns-74} M. J. Ablowitz, D. J. Kaup, A. C. Newell, H. Segur, Stud. Appl. Math., 53 (1974) 249.

\bibitem{gp-87} A.~V.~Gurevich and L.~P.~Pitaevskii, Zh. Eksp. Teor. Fiz.,  93 (1987)  871
[Sov. Phys.-JETP, 93 (1987) 871].

\bibitem{kamch-21a} A.~M.~Kamchatnov, Usp. Fiz. Nauk., { 191} (2021) 52
[Phys.--Uspekhi, {64} (2021) 48].

\bibitem{kamch-20a} A.~M.~Kamchatnov, 
Chaos { 30} (2020) 123148.

\bibitem{egkkk-07} G. A. El, A. Gammal, E. G. Khamis, R. A. Kraenkel, A. M. Kamchatnov,
Phys. Rev. A { 76} (2007) 053813.

\bibitem{egs-08} G.~A.~El, R.~H.~J.~Grimshaw,  N.~F.~Smyth, { Physica D}, { 237} (2008) 2423.

\bibitem{mfweh-20} M. D. Maiden, N. A. Franco, E. G. Webb, G. A. El, and M. A. Hoefer,
J. Fluid Mech. { 883} (2020) A10.

\bibitem{kamch-21} A.~M.~Kamchatnov, Zh. Eksp. Teor. Fiz. { 159} (2021) 76
[JETP, { 132} (2021) 63].

\bibitem{cbk-21} L.~F.~Calazans de Brito, A.~M.~Kamchatnov, 
Phys. Rev. E { 104} (2021) 054203.

\bibitem{ak-02} A. M. Kamchatnov and R. A. Kraenkel, J. Phys. A: Math. Gen., { 35} (2002) L13.

\bibitem{kamch-94} A. M. Kamchatnov, Phys. Lett. A, { 186} (1994)  387.

\bibitem{kamch-04} A. M. Kamchatnov, Physica D, { 188} (2004)  247.

\bibitem{ak-01} A. M. Kamchatnov, J. Phys. A: Math. Gen., { 34} (2001) L441.

\bibitem{krichever-88} I. M. Krichever, Funk. Analiz. Prilozh. { 22} (1988) 37
[Funct. Anal. Appl., { 22} (1988) 200].

\bibitem{karpman-67} V.~I.~Karpman, { Phys. Lett. A,} { 25} (1967) 708.

\bibitem{karpman-73} V. I. Karpman, {\it Non-Linear Waves in Dispersive Media,} (Nauka, Moscow, 1973)
(English translation: Pergamon Press, Oxford, 1975).

\bibitem{JLML-99} S. Jin, C. D. Levermore, D. W. McLaughlin,
Comm. Pure Appl. Math., { 52} (1999) 613.

\bibitem{kku-02} A. M. Kamchatnov, R. A. Kraenkel, B. A. Umarov,
Phys. Rev. E { 66} (2002) 036609.

\bibitem{bouss-1877}  J. Boussinesq, 
M\'{e}m. Pr\'{e}s. Div. Sav. Acad. Sci. Inst. Fr. 23 (1877) 1.

\bibitem{kaup-75}  D. J. Kaup, 
Prog. Theor. Phys. 54 (1975) 396.

\bibitem{ikcp-17} S. K. Ivanov, A. M. Kamchatnov, T. Congy, and N. Pavloff,
Phys. Rev. E { 96} (2017)  062202.

\bibitem{kbhp-88} C. F. Kennel, B. Buti, T. Hada, and R. Pellat, Phys. Fluids, 31 (1988) 1949.

\bibitem{kn-78} D. J. Kaup and A. C. Newell, 
J. Math. Phys. 19 (1978) 798.

\bibitem{ckp-16} T. Congy, A. M. Kamchatnov, and N. Pavloff,
SciPost Phys., 1 (2016) 006.

\bibitem{ish-17} E. Iacocca, T. J. Silva, and M. A. Hoefer, 
Phys. Rev. Lett. { 118} (2017) 017203.

\bibitem{br-81} A. E. Borovik and V. N. Robuk,  Teor. Mat. Fiz. 46 (1981) 371 [Theor. Math. Phys. 46 (1981) 242].

\bibitem{ks-23} A. M. Kamchatnov and D. V. Shaykin, Propagation of generalized Korteweg-de Vries solitons 
along large scale waves, preprint arxiv:2307.12321 (2023).



\end{thebibliography}
\end{document}